\documentclass[12pt]{article}
\usepackage{amssymb, amsthm}
\usepackage{amsmath,amsfonts}
\usepackage{graphicx}
\usepackage{booktabs}
\usepackage{fullpage}
\usepackage{natbib}
\usepackage{url} 
\usepackage{color,verbatim}
\usepackage{tikz}
\usetikzlibrary{matrix,chains,positioning,decorations.pathreplacing,arrows}

\usepackage{algorithmic,algorithm}
\usepackage{changes}

\theoremstyle{plain}

\theoremstyle{definition} 

\begin{document}
\begin{center}
{\bf\large Application of Kriging Models for a Drug Combination Experiment on Lung Cancer}

{Qian Xiao$^1$, Lin Wang$^2$, and Hongquan Xu$^2$}

$^1$Department of Statistics, University of Georgia, Athens, Georgia 30602, U.S.A. \\
$^2$Department of Statistics, University of California, Los Angeles, California 90095, U.S.A.

(\today)
\end{center}

\begin{quote}{\em Abstract:}
{
Combinatorial drugs have been widely applied in disease treatment, especially chemotherapy for cancer, due to its improved efficacy and reduced toxicity compared with individual drugs. The study of combinatorial drugs requires efficient experimental designs and proper follow-up statistical modelling techniques. Linear and non-linear models are often used in the response surface modelling for such experiments. We propose the use of  Kriging models to better depict the response surfaces of combinatorial drugs and take into account the measurement error. We further study how proper experimental designs can reduce the required number of runs.  We illustrate our method via a combinatorial drug experiment on lung cancer. We demonstrate that only 27 runs are needed to predict all 512 runs in the original experiment and achieve better precision than existing analysis.
}
\end{quote}

\begin{quote}
\end{quote}

\section{Introduction}
 \label{intro}

Combination chemotherapy with multiple drugs have been widely applied in cancer therapy. Such combinatorial drugs have enhanced efficacy and reduced toxicity due to multiple targets and synergistic drug interactions (\cite{devita1975}, \cite{lilenbaum2005} and \cite{ning2014}). Preclinical experiments in vitro are usually conducted to characterize the  pathological mechanisms and find the optimal drug combinations. In the analysis of these experiments, different response surface modeling techniques are used to quantify the dose-effect relationships. For economic reasons, the response surface model that requires less runs and has better predictive power at the same time are preferred. 

For combinatorial drugs with only two components, Hill models based on ray designs (\cite{chou2006}) are popular in the analysis, but they are not suitable for multiple drug combinations (\cite{zhou2014}). Polynomial (linear) models accompanied by full factorial or fractional factorial designs are often used in analyzing multiple drug combinations (\cite{jaynes2013}), but their outputs are not bounded. In practice, many such experiments require bounded responses, e.g. survival rate.  
The Hill-based (non-linear) model (\cite{ning2014}) is a combination of Hill models and polynomial models, which overcome the shortcomings of both. But, it is not stable in many situations. Neural networks can also be applied  (\cite{al2011}), but it require many data and the interpretations are hard. 

In this paper, we propose the use of Kriging model, and show its  superiority compared with other modeling techniques in a combinatorial drug experiment on lung cancer. In this experiment conducted by \cite{al2011}, a 512-run, 8-level and 3-factor full factorial design was applied to both normal cells and lung cancer cells. \cite{ning2014} analyzed the same experiment with the Hill-based model. 
In this paper, we show that the Kriging model always performs the best dealing with various data sizes. Moreover, it can use only 27 runs to predict all 512 runs with the highest accuracy, compared with the polynomial model and Hill-based model using all 512 runs in \cite{ning2014} and the neural network using the recommended 80 runs in \cite{al2011}. Kriging models are robust under various experimental designs and can efficiently identify the drug interactions. Sensitivity analysis can be done to select significant factors, and measurement errors are taken into consideration in Kriging models.


This paper is organized as follows. In Section \ref{model}, we introduce four major response surface modeling techniques. We illustrate Kriging models in details, and show the neural network  used in \cite{al2011} and the polynomial and Hill-based models used in \cite{ning2014}. 
In Section \ref{raan}, we compare these four models in analyzing the combinatorial drug experiment on lung cancer which are used in both \cite{al2011} and  \cite{ning2014}. Section \ref{concl1} concludes and discusses some future research. 

\section{Response surface modeling}
\label{model}


\subsection{The Kriging model}
\label{skm}

Originally from geosciences (\cite{krige1951}), Kriging models are now widely used in computer experiments for optimization and sensitivity analysis. Computer experiments are popular in scientific researches and product developments to simulate real-world problems with complex and deterministic computer codes.  Kriging models can compensate for the effects of data clustering and give better estimation of prediction error. In a Kriging model, the responses are viewed as realizations of a Gaussian process, and the predicted response at a target point can be represented as a weighted average of the responses at observed points. 
For an introduction to Kriging models, see \cite{sacks1989},  \cite{kleijnen2009},  \cite{ginsbourger2009} and \cite{cressie2015statistics}. 

\begin{figure}
    \centering
    \caption{Examples of spatial correlation functions of Mat\'ern family }
    \includegraphics[width=0.4\linewidth]{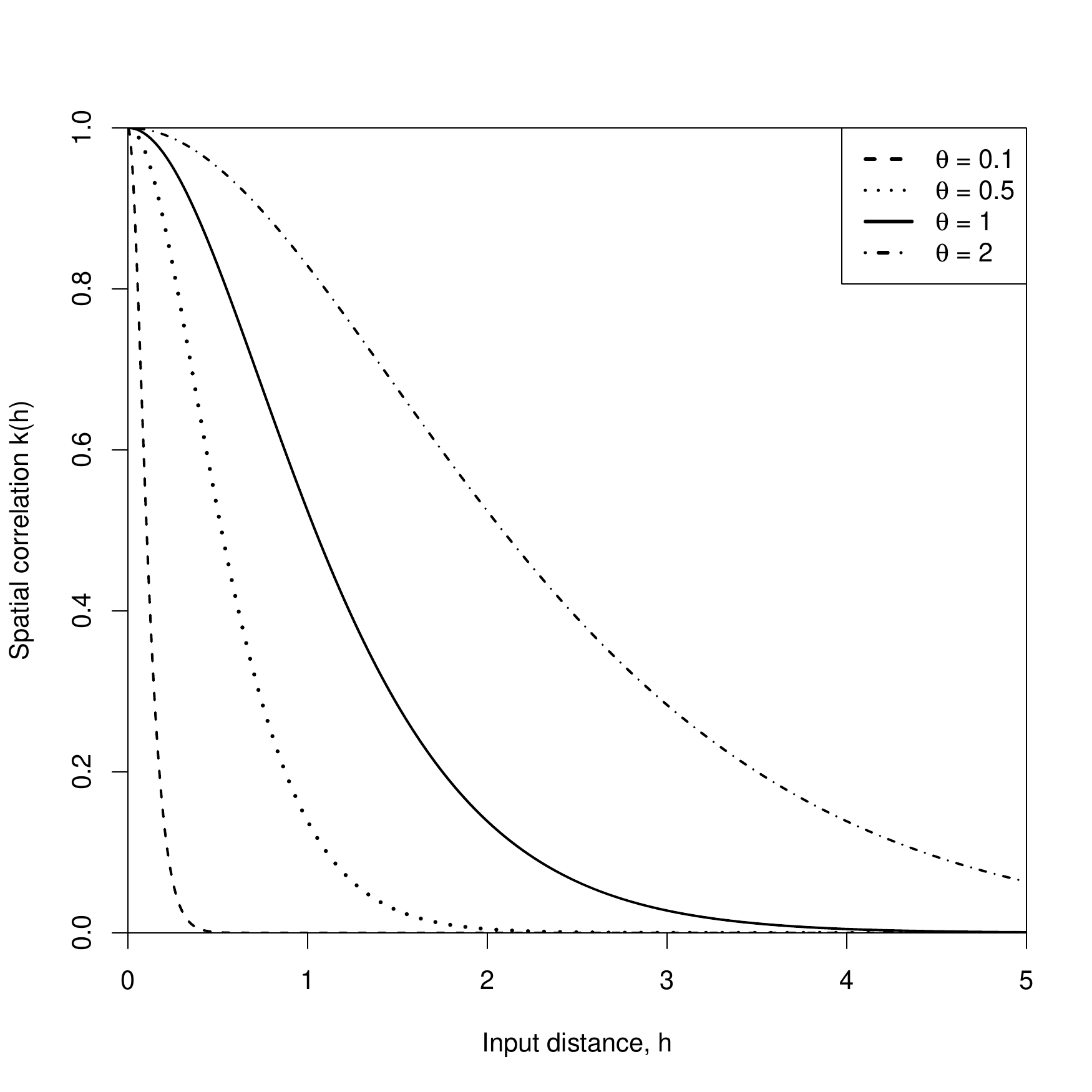}
    \label{fig:rp}
\end{figure}

Different from the deterministic case in computer experiments, Kriging for random simulations should be used in combinatorial drug experiments due to the existence of measurement errors. It is desirable to adopt the following Ordinary Kriging (OK) model with a noise term
\begin{equation}
\label{kgm}
y(x) = \mu +Z(x) + \epsilon,
\end{equation}
where $y(x)$ is the response at point $x$,  $\mu$ is the trend (or intercept), $Z(x)$ is a Gaussian process with zero mean and  constant variance, and $\epsilon(x) \sim N(0, \tau^2)$ is independent of $Z(x)$. The covariance function for $Z(x)$ is defined as:
\begin{equation}
\phi(x_i, x_j) = cov(Z(x_i), Z(x_j)) =  \sigma^2 *\prod_{l=1}^{d} K_l(h),
\label{eqcov}
\end{equation}
where $h = \vert x_{i,l}-x_{j,l}\vert$, $x_{i,l}$ and $x_{j,l}$ are the $l^{th}$ elements of points (runs) $x_i$ and 
$x_j$, $d$ is the dimensions (number of factors), $\sigma^2$ is the variance parameter, and $K_l(h)$ is the chosen stationary correlation function. Two  popular types of $K(h)$ are:
$$
\text{Gauss: \ \ }  K_l(h)= exp(-1/2*(h/\theta_l)^2),
$$
$$
\text{Mat\'ern} (\nu = p+1/2): \ \  
K_l(h)= exp\left(-\frac{\sqrt{2\nu}h}{\theta_l}\right)
\frac{\Gamma(p+1)}{\Gamma(2p+1)}
\sum_{i=0}^p\frac{(p+i)!}{i!(p-i!)}\left(\frac{\sqrt{8\nu}h}{\theta_l}\right)^{p-i},
$$
where $p \in \mathbb{N}^+$, $\theta_l$ is the range parameter which scales the correlation length, and $\Gamma()$ is the gamma function. The sample paths of $z(x)$ with Gaussian correlation have derivatives at all orders and are too smooth, which may cause numeric problems. \cite{rasmussen2006} and \cite{martin2005} recommended the use of Mat\'ern correlation with $\vartheta = 5/2$  where
$
K_l(h)= (1+{\sqrt{5}h} / {\theta_l}+{5} / {3}*({h} / {\theta_l})^2)
*exp(-{\sqrt{5}h} / {\theta_l})
$, 
and $z(x)$ is twice differentiable.
Figure \ref{fig:rp} shows the Mat\'ern correlations ($\vartheta = 5/2$) with different range parameters $\theta$. With smaller $\theta$, the correlation decreases faster to zero as $h$ increases.
The parameters $\theta_l$ ($l=1, \ldots,d$) and $\sigma^2$ in the correlation function can be estimated by maximizing a likelihood function (MLE) based on the observed data.
With the estimated parameters, the best linear unbiased prediction on any point $x$ is
\begin{equation}
\label{pred}
\widehat{y}(x) = \widehat{\mu} + \gamma^T C^{-1}(\textbf{y} - \widehat{\mu} \textbf{1}),
\end{equation}
where $\textbf{y}$ is the response vector at $n$ observed points $x_1, \ldots, x_n$,
$
\widehat{\mu} = (\textbf{1}^{T}C^{-1}\textbf{1})^{-1}\textbf{1}^{T}C^{-1}\textbf{y}
$ is an estimate of $\mu$,
$\gamma$ is the covariance vector $(\phi(x,x_1), \ldots, \phi(x,x_n))^{T}$, matrix $C = \Phi + \Delta$, $\Phi$ is the variance-covariance matrix $(\phi(x_i,x_j))_{1\leqslant i,j\leqslant n}$, $\Delta$ is a diagonal matrix with diagonal elements $\tau^2$, and $\textbf{1}$ is a column of $n$ ones. From Equation \ref{pred}, it's easy to show that this model does not interpolate all observed data due to the existence of measurement errors ($\tau^2 \neq 0$). 
In the unreplicated experiment studied in Section \ref{raan}, we assume a homogeneous variance $\tau^2 = 0.0001$ for the noise term, since the measurement is roughly accurate to 2 decimal places. Choosing $\tau^2$ within the range 0.001 to 0.00001 does not make significant difference in the model estimation. We adopt the R package ``DiceKriging" (\cite{roustant2009}) to estimate the Kriging models in this paper. 

\subsection{Neural networks}

Neural networks (\cite{mcculloch1943}) are widely used in machine learning, pattern recognition, medical diagnosis and many other areas.
An (artificial) neural network is based on a collection of connected units called neurons, which receive input and produce output via its network function. 
Neural network models are very flexible and it is generally hard to determine the best network structures in practice.  For a detailed introduction to neural networks, see \cite{livingstone2008}.

\pagestyle{empty}
\def\layersep{2.5cm}
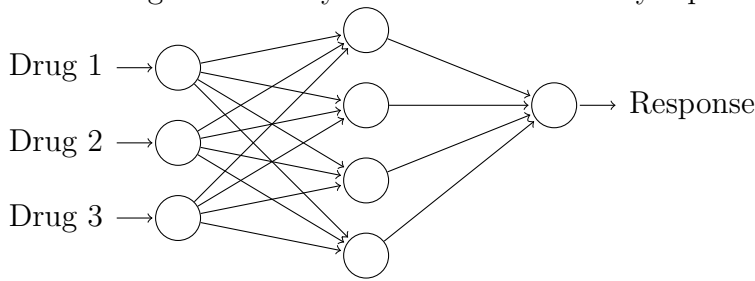
\begin{figure}
    \centering
    \caption{A single hidden layer four-neuron multilayer perceptron}
    
    \begin{tikzpicture}[shorten >=1pt,->,draw=black!100, node distance=\layersep]
    \tikzstyle{every pin edge}=[<-,shorten <=1pt]
    \tikzstyle{neuron}=[circle,fill=black!25,minimum size=17pt,inner sep=0pt]
    \tikzstyle{input neuron}=[neuron, fill=white!100,draw=black];
    \tikzstyle{output neuron}=[neuron, fill=white!100,draw=black];
    \tikzstyle{hidden neuron}=[neuron, fill=white!100,draw=black];
    \tikzstyle{annot} = [text width=4em, text centered]

\foreach \name / \y in {1,...,3}
    \node[input neuron, pin=left:Drug \y] (I-\name) at (0,-\y) {};

\foreach \name / \y in {1,...,4}
    \path[yshift=0.5cm]
        node[hidden neuron] (H-\name) at (\layersep,-\y cm) {};

\node[output neuron,pin={[pin edge={->}]right:Response}, right of=H-2] (O1) {};
\foreach \source in {1,...,3}
    \foreach \dest in {1,...,4}
        \path (I-\source) edge (H-\dest);

\foreach \source in {1,...,4}
    \path (H-\source) edge (O1);
\end{tikzpicture}
    \label{fig:nn}
    
Note: the bias nodes (with value 1) are not shown in this figure.
\end{figure}

\cite{al2011} fitted a multilayer perceptron (shown in Figure \ref{fig:nn}) in analyzing the combinatorial drug experiment.  
In this model, for the $j^{th}$ hidden neuron ($j=1,2,3,4$), the network function is
$$
f(x) = \frac{1}{1+e^{-\sum w_{i,j} x_{i,j}}},
$$
where $w_{i,j}$ are parameters to be estimated, $x_{0,j} = 1$ and $x_{i,j}$ is the $i^{th}$ input value ($i = 1, 2, 3$). For the output neuron, $
g(x) = \sum w^{'}_j x^{'}_j
$ where $w_j^{'}$ are parameters to be estimated, $x^{'}_0 = 1$ and $x_j^{'}$ is the output from the $j^{th}$ hidden neuron ($j = 1, 2, 3,4$).  Neural networks can be estimated via resilient back-propagation method, and R package "neuralnet" (\cite{fritsch2016}) is a current popular tool. 

\subsection{Polynomial and Hill-based models}
Polynomial models are the most common analytic tools in drug experiments. In both \cite{al2011} and \cite{ning2014},   the polynomial model in Equation \ref{lmr}  is used, which includes all main, interaction and quadratic effects for the three drugs A,B and C.

\begin{equation}
\label{lmr}
y = \beta_0 + \beta_1A + \beta_2B +\beta_3C +\beta_4AB+ \beta_5AC + \beta_6BC + \beta_7A^2 + \beta_8B^2 + \beta_{9}C^2 + \epsilon \ . 
\end{equation}

In vivo system, the dosage-effect relationship usually follows a sigmoidal curve (\cite{chou2006}). Based on this, \cite{ning2014} combined the polynomial and Hill models, and proposed the Hill-based model:
\begin{equation}
\label{hilr}
y = \frac{1}{1+(\frac{c}{IC_{50}(\theta)})^{\gamma(\theta)}} + \epsilon\ ,
\end{equation}
where the total dosage $c = c_1+c_2+c_3$, and $c_1$, $c_2$ and $c_3$ are the actual dosages of drugs A, B and C, respectively;  the dosage proportion $\theta_i = c_i/c$ ($i = 1, 2, 3$); $IC_{50}(\theta) = a_0 +a_1\theta_1 +a_2\theta_2 + a_3\theta_1\theta_2 + a_4\theta_1^2 +a_5\theta_2^2$; $\gamma(\theta) = b_0 +b_1\theta_1 +b_2\theta_2 + b_3\theta_1\theta_2 + b_4\theta_1^2 +b_5\theta_2^2$.  Function $IC_{50}(\theta)$ measures the dosage of the drug combination which yields $50\%$ effect level, and $\gamma(\theta)$ measures the changing rate of the smooth curve. Hill-based models are able to address
all drug combinations and characterize the interaction patterns. The responses from Hill-based models are bounded within range 0 to 1.

\section{Results and analysis}
\label{raan}

\subsection{Drug combination experiment on Lung cancer}

In this section, we focus on the drug combination experiment on lung cancer conducted by \cite{al2011}.
Three drugs AG490(A), U0126(B) and indirubin-$3^{'}$-monoxime (I-3-M)(C) which are inhibitors targeting signaling pathways for cell survival and proliferation were used, and a 512-run, 8-level full factorial design ($D_{\text{full}}$) was applied to both normal cells and lung cancer cells. The response variable is the ATP level (standardized to 0-1 range) of the cell measured 72 hours after the drug treatments. The actual dosages for each drug are given in Table \ref{drugtab1} and coded as level 0 to 7. The purpose of this experiment is to model the response surface and systematically quantify the characterization of
cellular responses. An optimal combinatorial drug in this study should minimize the ATP levels of cancer cells while keeping the ATP levels of normal cells above certain standards.

\begin{table}  
\caption{Dose levels for each drug in the combinatorial experiment on lung cancer}
\label{drugtab1}
\centering
    \begin{tabular}{|c|r|r|r|r|r|r|r|r|}
    \hline
    Drug  & \multicolumn{8}{c|}{Dosage ($\mu$M)}\\
    \hline
    AG490 (A) & 0     & 0.3   & 1     & 3     & 10    & 30    & 100   & 300 \\
    \hline
    U0126 (B) & 0     & 0.1   & 0.3   & 1     & 3     & 10    & 30    & 100 \\
    \hline
    I-3-M (C)  & 0     & 0.3   & 1     & 3     & 10    & 30    & 100   & 300 \\
    \hline
    Coded level  & 0     & 1   & 2     & 3     & 4    & 5    & 6   & 7 \\
    \hline
    \end{tabular}%
\end{table}%

\subsection{Model fitting and  comparison}

When comparing Kriging models, neural networks, polynomial models and Hill-based models, we consider four possible designs: the original 512-run 8-level full factorial design $D_{\text{full}}$, a 80-run random sub-design $RD_{80}$, a 27-run random sub-design  $RD_{27}$, and a 27-run, 3-level (coded levels 0, 4, 7 in Table \ref{drugtab1}) full factorial design $D047$. We use the actual dosages (standardized to 0-1 range) in all these design matrices. Given an $n$-run design, we fit a model using $n$ observations and use the model to predict all 512 observations. Then we compute the mean square error (MSE) and correlations ($r$) based on the 512 predicted and actual responses. 

Tables \ref{tab:nor} and \ref{tab:can} compares $1000 \times MSE (r)$ from different models and designs for normal and cancer cells, respectively. ``Neural network" is a single-layer four-neuron neural network of which the result varies slightly each time running the R package "neuralnet".  we select the best result among 100 repetitions.
Results for designs $RD_{80}$ and $RD_{27}$ are average values from 100 random designs. When fitting Hill-based models with either $RD_{80}$ or $RD_{27}$, numeric problems may occur, and we exclude them when calculating the average.

\begin{table}[htbp]
\caption{Comparison of models and designs in fitting normal cell data}
\begin{center}
    \begin{tabular}{lllll}
    \hline
          & $D_{\text{full}}$ & $RD_{80}$ & $RD_{27}$ & $D047$ \\
          \hline
    Kriging & 0.0018 (100.00\%) & 0.21(99.88\%) & 0.97(99.56\%) & 0.31(99.86\%) \\
    Neural network   & 0.11(99.95\%) & 1.28(99.37\%) & 3.52(98.43\%) & 2.99(98.61\%)  \\
    Polynomial & 0.48(99.75\%) & 1.16(99.42\%) & 3.65(98.39\%) & 1.12(99.49\%) \\
    Hill-based & 0.89(99.10\%) & 1.07(99.49\%) & 3.57(98.30\%) & 3.30(98.39\%) \\
    \hline
    \end{tabular}%
\end{center}
  \label{tab:nor}%
\end{table}%

\begin{table}[htbp]
  \caption{Comparison of models and designs in fitting cancer cell data}
  \begin{center}
    \begin{tabular}{lllll}
    \hline
         & $D_{\text{full}}$ & $RD_{80}$ & $RD_{27}$ & $D047$ \\
          \hline
    Kriging & 0.0030 (100.00\%) & 0.37(99.78\%) & 1.84(99.23\%) & 1.05(99.65\%) \\
    Neural network   & 0.27(99.88\%) & 1.57(99.34\%) & 3.97(98.43\%) & 2.98(99.16\%) \\
    Polynomial & 2.98(98.67\%) & 6.77(97.09\%) & 39.82(87.74\%) & 5.84(97.66\%) \\
    Hill-based & 1.42(98.80\%) & 1.67(99.33\%) & 4.99(97.93\%) & 4.70(97.92\%) \\
    \hline
    \end{tabular}%
  \end{center}
  \label{tab:can}%
\end{table}%

From Tables \ref{tab:nor} and \ref{tab:can}, we can see that for both normal and cancer cells, Kriging models are always the best in prediction (smallest MSEs and largest  correlations) for all four types of designs. In addition, Kriging models have the least number of parameters, and are suitable for high dimension data. Note that when fitting Kriging, neural network, polynomial and Hill-based models in this experiment, the numbers of parameters to be estimated are 5, 21, 10 and 12, respectively. In Table \ref{tab:pp1}, we show the estimated parameters and their standard deviations (SDs) for  Kriging models along with designs $D_{\text{full}}$ and $D047$. 
All $\theta$ are significantly different from 0, thus there is no identifiability issues. The SDs are computed from 1000 simulations.

\begin{table}[htbp]
  \caption{Estimations of "parameters(SDs)" in Kriging models}
  \begin{center}
    \begin{tabular}{llllll}
    \hline
    Normal cells & $\theta_A$ & $\theta_B$ & $\theta_C$ & $\sigma^2$ & trend \\
    \hline
    $D_{\text{full}}$ & 1.24(0.11)  & 2.00(0.04)  & 1.24(0.12)  & 0.26(0.04)  & 0.62(0.06) \\
    $D047$ & 1.11(0.22)  & 1.89(0.26)  & 1.08(0.22)  & 0.24(0.05)  & 0.54(0.10) \\    
    \hline
    Cancer cells & $\theta_A$ & $\theta_B$ & $\theta_C$ & $\sigma^2$ & trend \\
    \hline
    $D_{\text{full}}$ & 0.98(0.11)  & 1.21(0.13)  & 0.52(0.06)  & 0.12(0.04)  & 0.39(0.04) \\
    $D047$ & 0.83(0.20)  & 1.46(0.23)  & 0.41(0.10)  & 0.16(0.04)  & 0.37(0.06)\\
    \hline
    \end{tabular}%
  \end{center}
  \label{tab:pp1}%
\end{table}%

Using Kriging model, a small design can be sufficient in depicting the response surface. From Tables \ref{tab:nor} and \ref{tab:can}, we can see that when using Kriging models and 27-run design $D047$, the MSEs are  as small as $3.10*10^{-4}$ for normal cells and $1.05*10^{-3}$ for cancer cells. As a comparison, when using Hill-based models and 512-run design $D_{\text{full}}$, the MSEs are $8.91*10^{-4}$ and $1.42*10^{-3}$; when using polynomial models and  $D_{\text{full}}$, the MSEs are $4.8*10^{-4}$ and $2.98*10^{-3}$; when using neural networks and 80-run design $RD_{80}$, the MSEs are $1.28*10^{-3}$ and $1.57*10^{-3}$, for normal and cancer cells, respectively. It's clear that Kriging models require the least number of runs and give the best predictions. 
The structured design $D047$ outperforms the random design $RD_{27}$, and is good enough in prediction under Kriging models. In addition, design $D047$ is robust for all four types of models; while, designs $RD_{80}$ and $RD_{27}$ are unstable. When fitting Hill-based models with random 100 designs of $RD_{80}$ and $RD_{27}$, numeric problems occurred 6 and 35 times, respectively.

\begin{figure}[htp]
  \centering
  \caption{Scatter-plots of predicted versus observed ATP levels on normal cells using design $D047$.} 
  \begin{tabular}{cc}


    \includegraphics[width=0.45\linewidth]{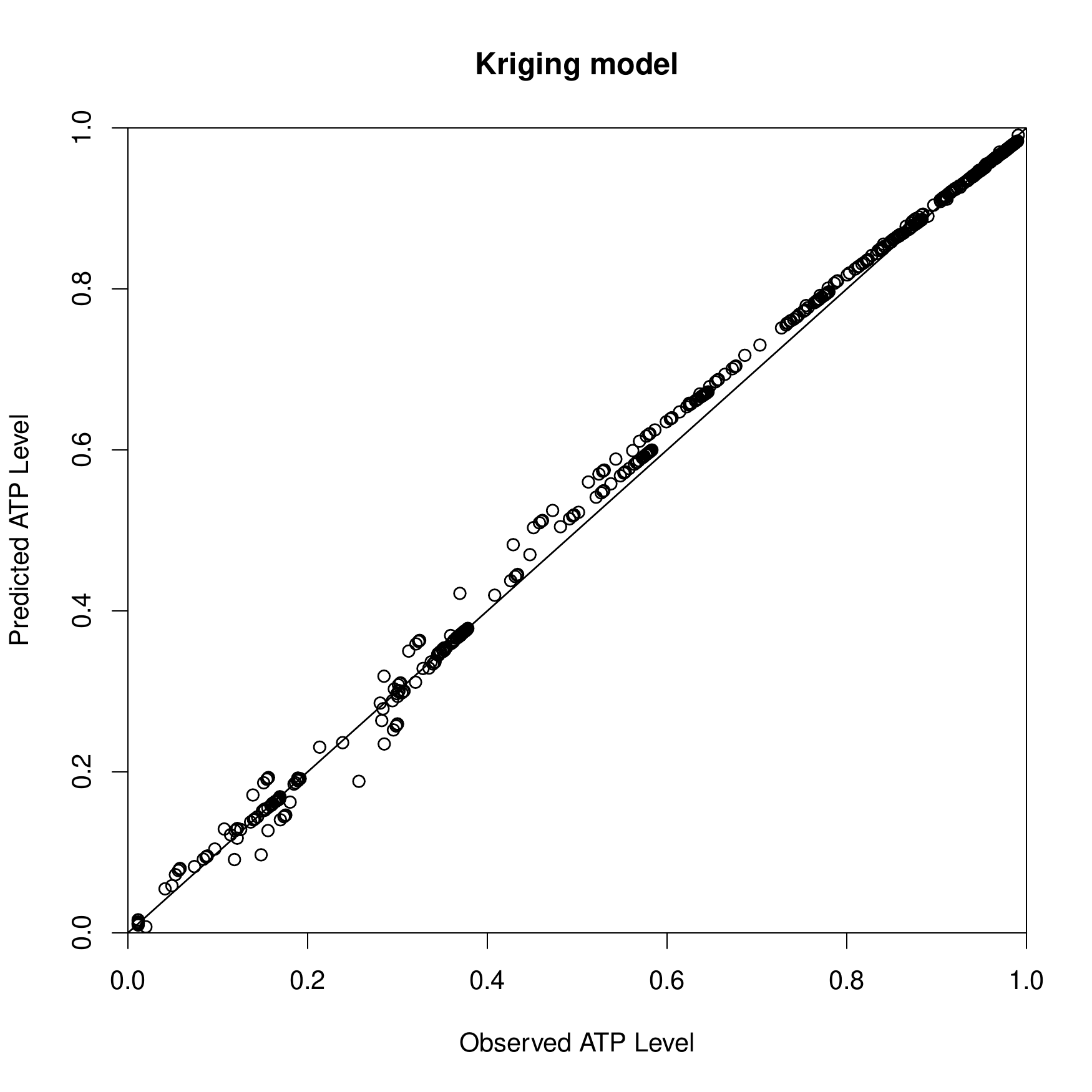}&

    \includegraphics[width=0.45\linewidth]{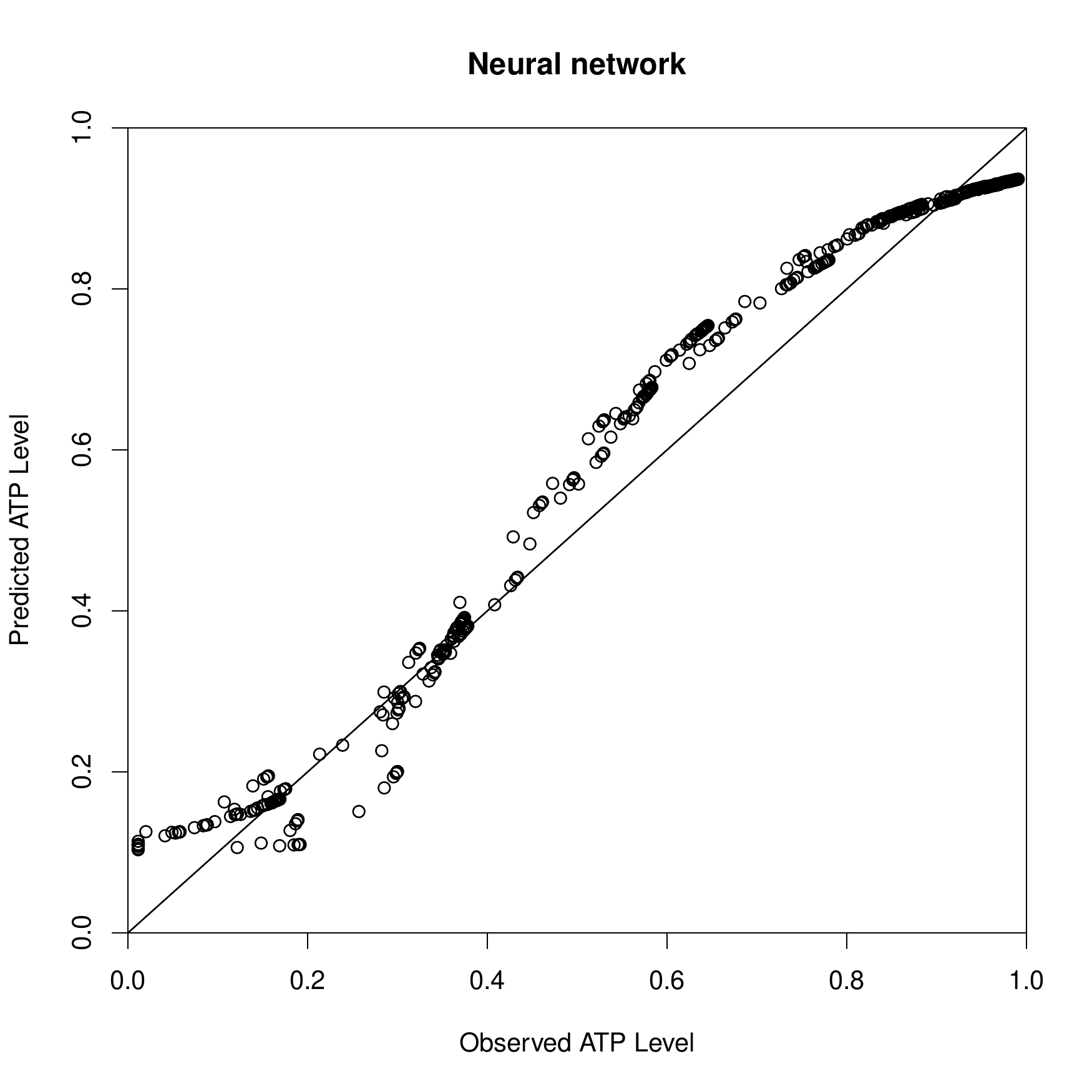}\\

    \includegraphics[width=0.45\linewidth]{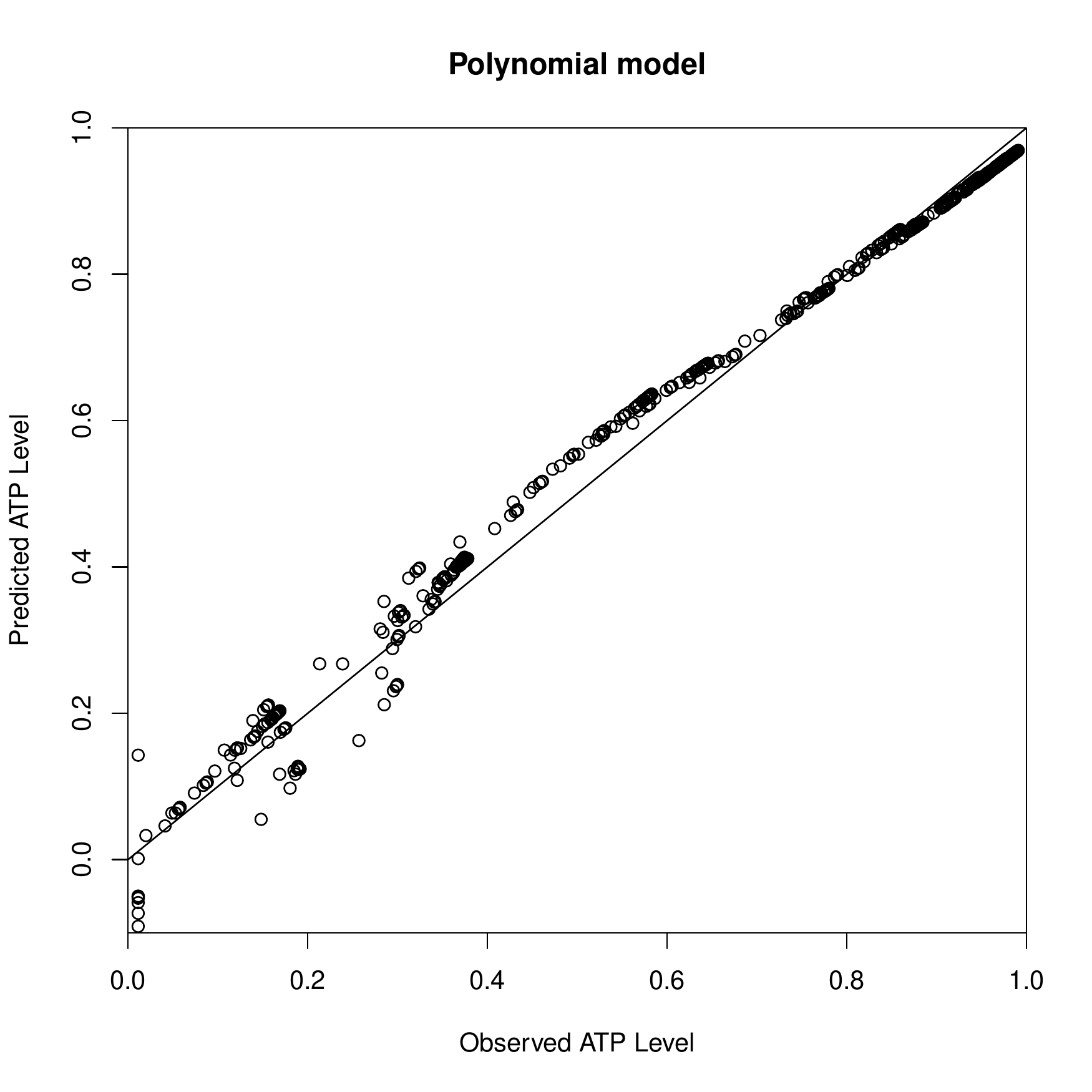}&

    \includegraphics[width=0.45\linewidth]{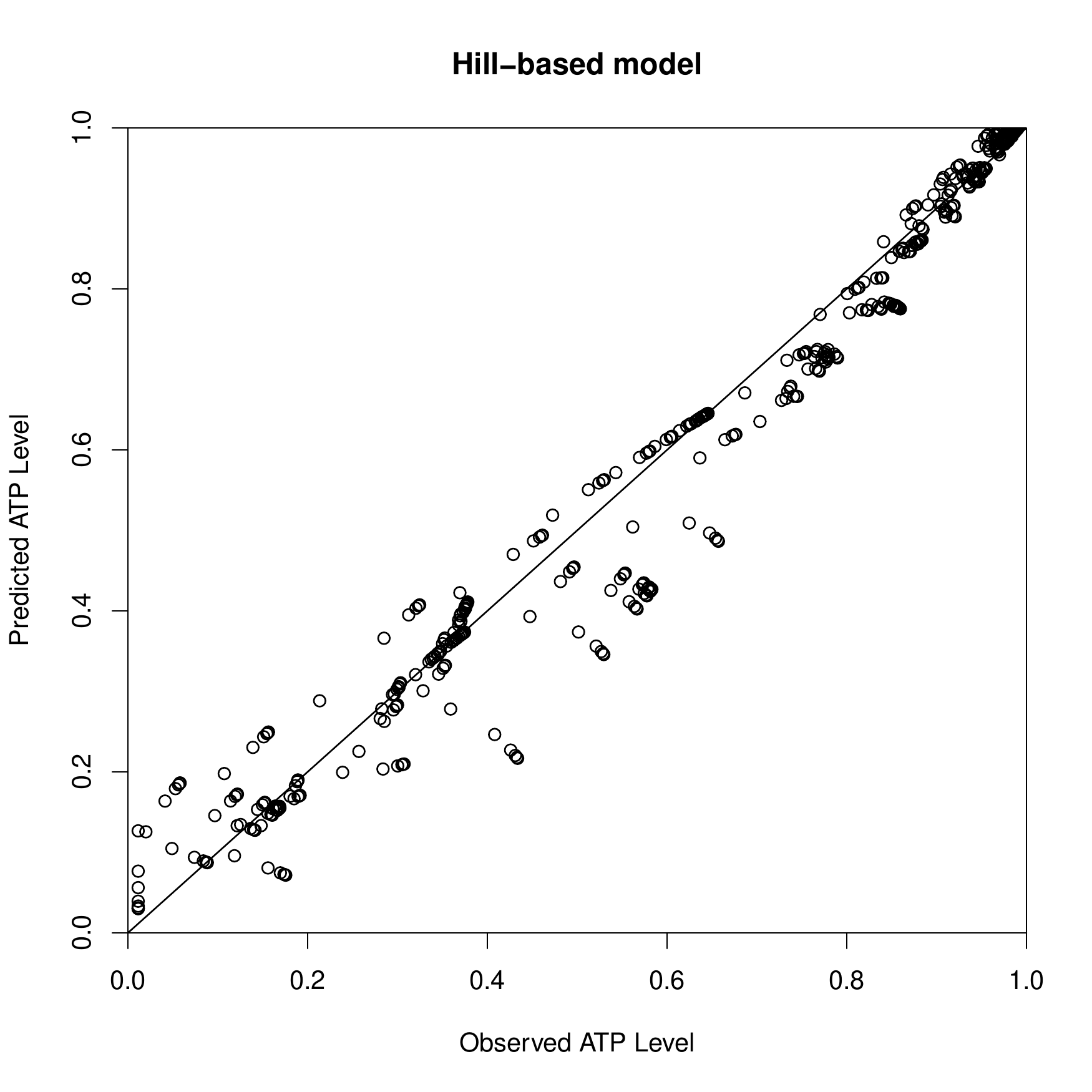}\\
  \end{tabular}
    \label{fig:nor}
\end{figure}

\begin{figure}[htp]
  \centering
  \caption{Scatter-plots of predicted versus observed ATP levels on cancer cells using design $D047$.}
  \begin{tabular}{cc}


    \includegraphics[width=0.45\linewidth]{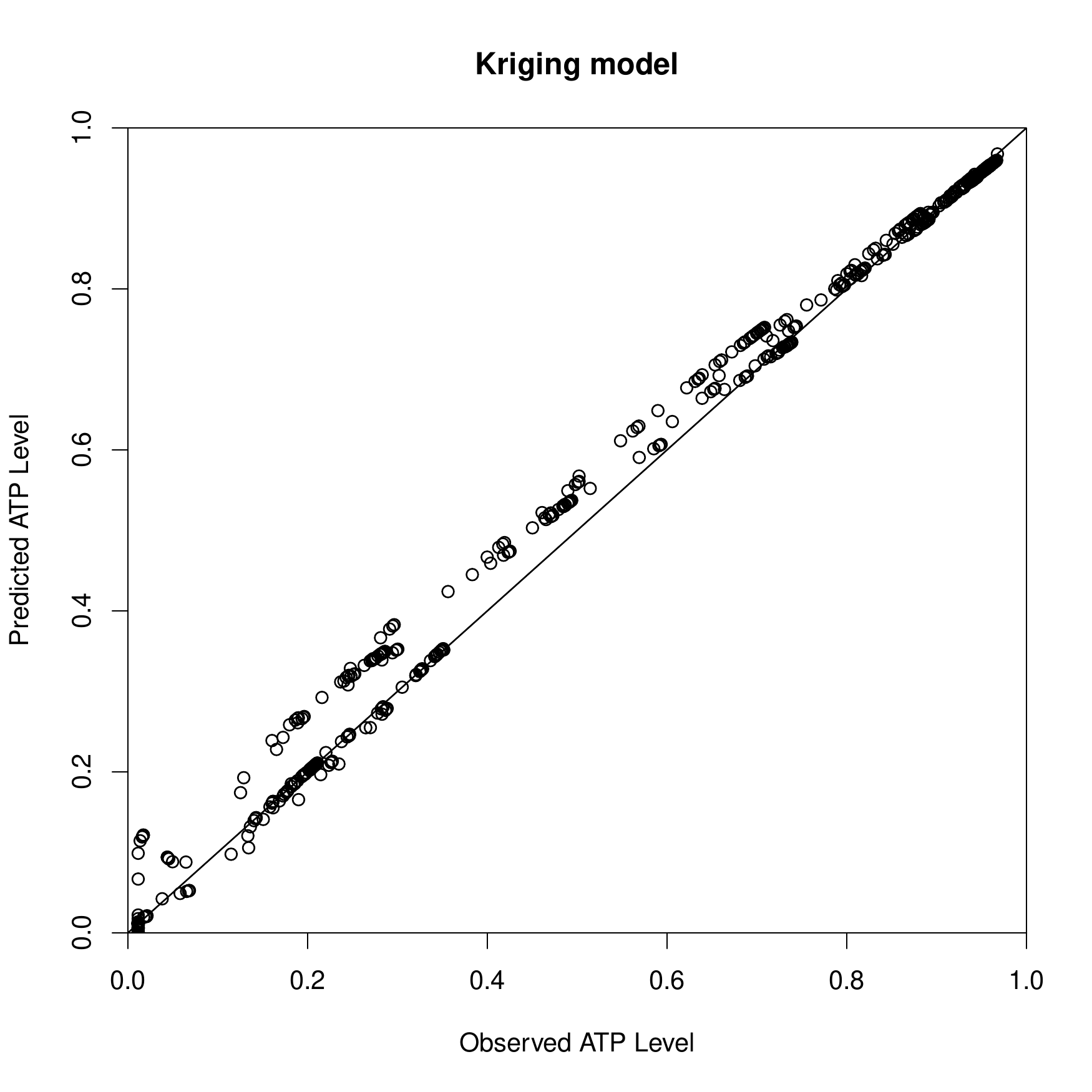}&

    \includegraphics[width=0.45\linewidth]{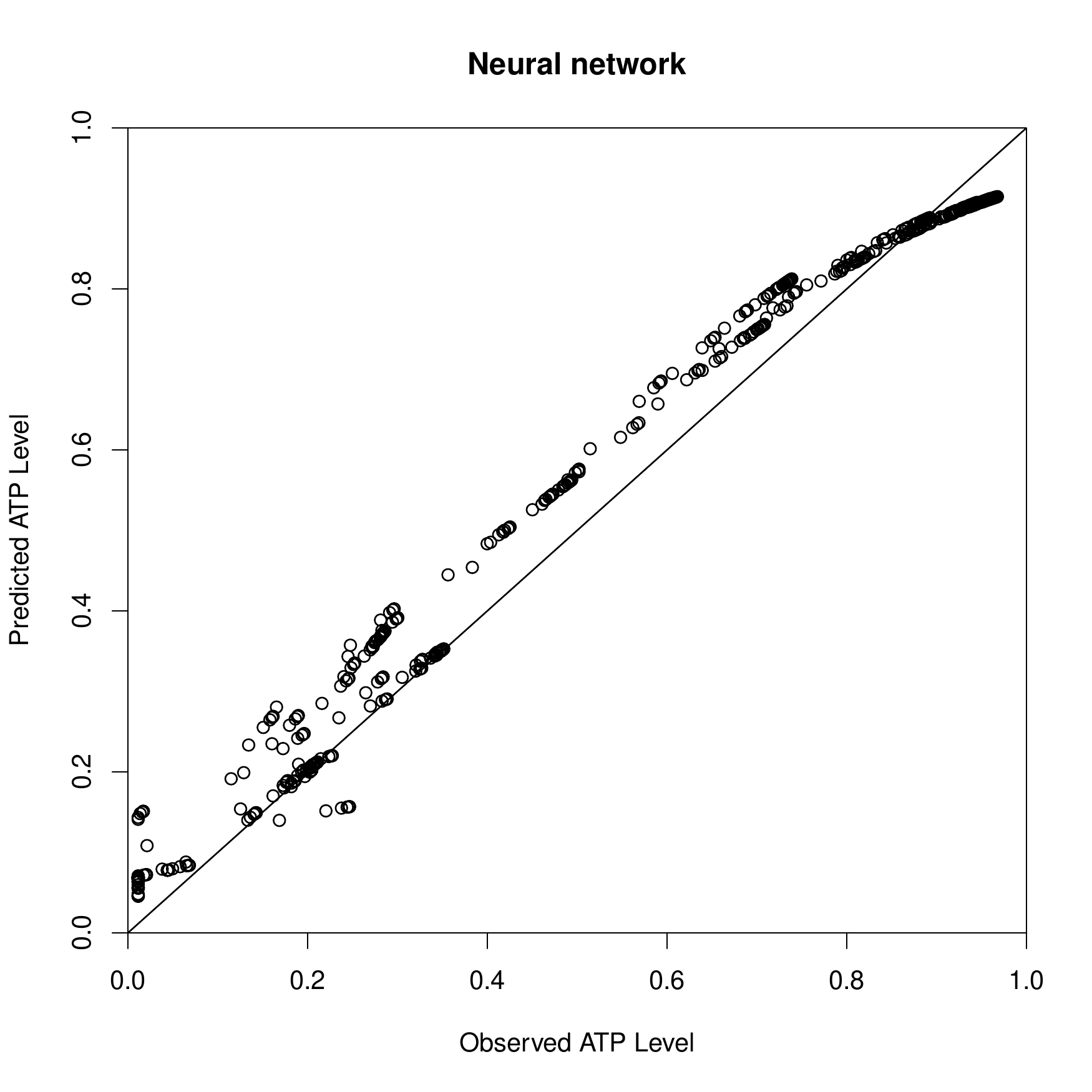}\\

    \includegraphics[width=0.45\linewidth]{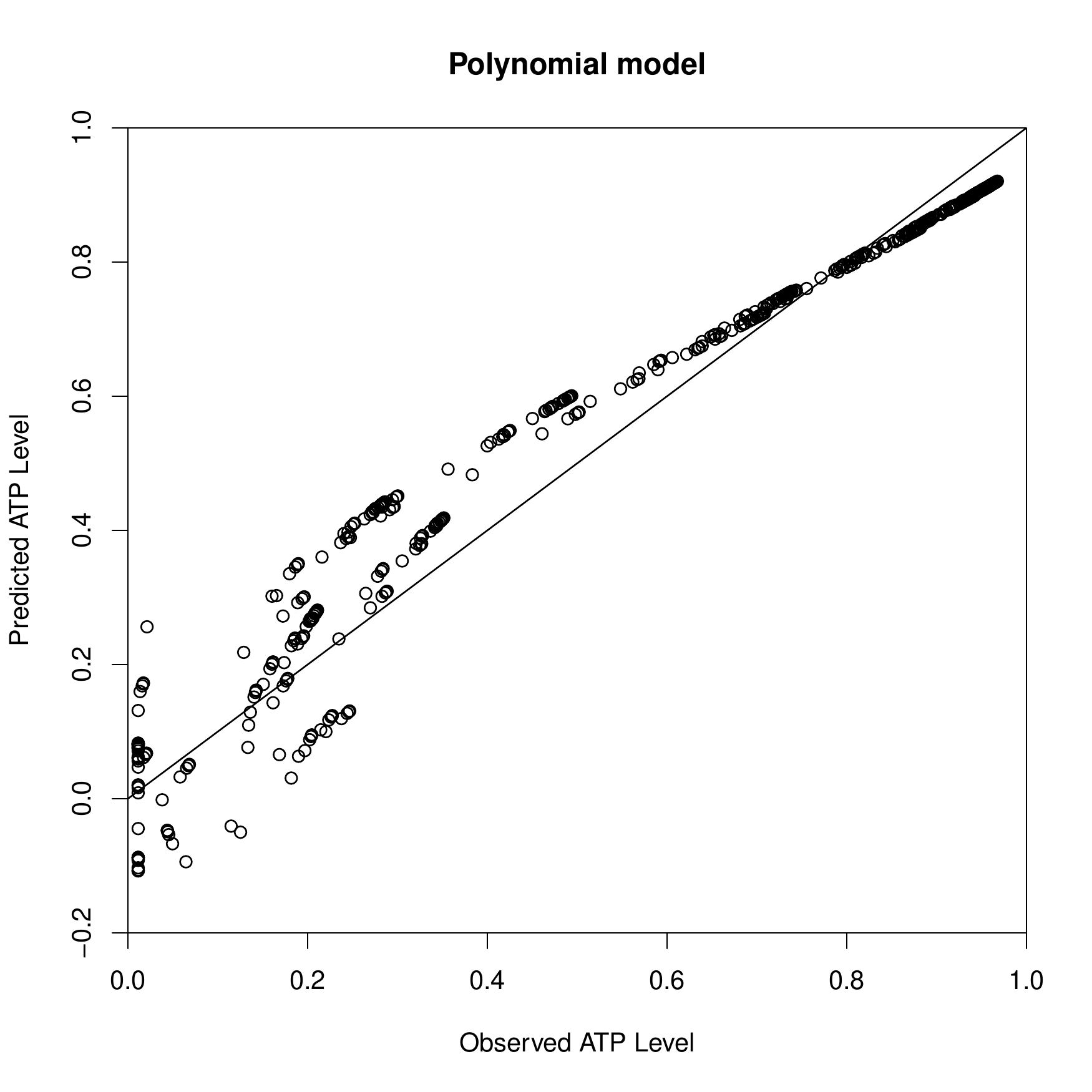}&

    \includegraphics[width=0.45\linewidth]{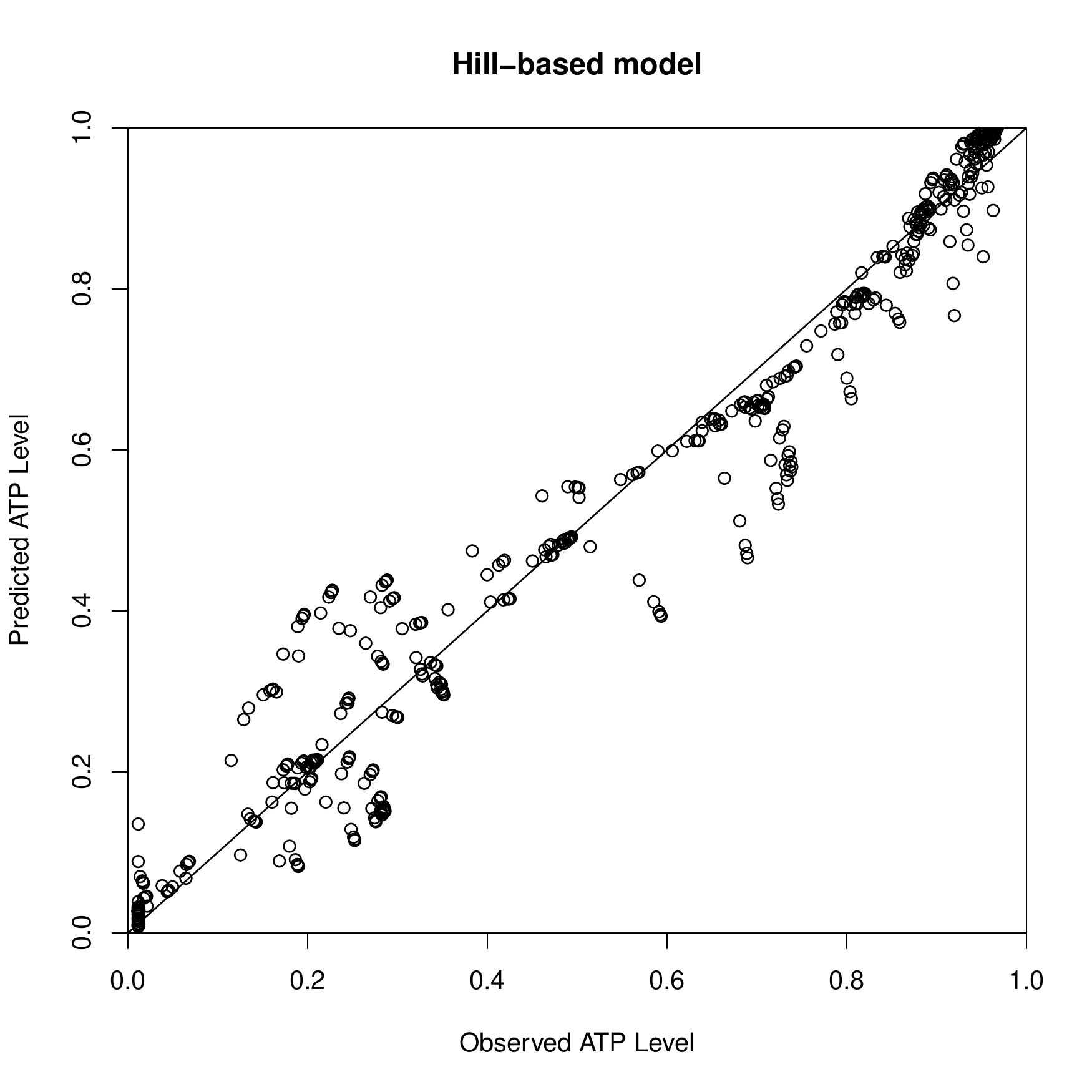}\\
  \end{tabular}
    \label{fig:can}
\end{figure}

\begin{figure}
\caption{Contour plots of predicted ATP levels via $D_{\text{full}}$ and $D047$ under Kriging models on  normal cells.}
    \centering
    \includegraphics[width=0.8\linewidth]{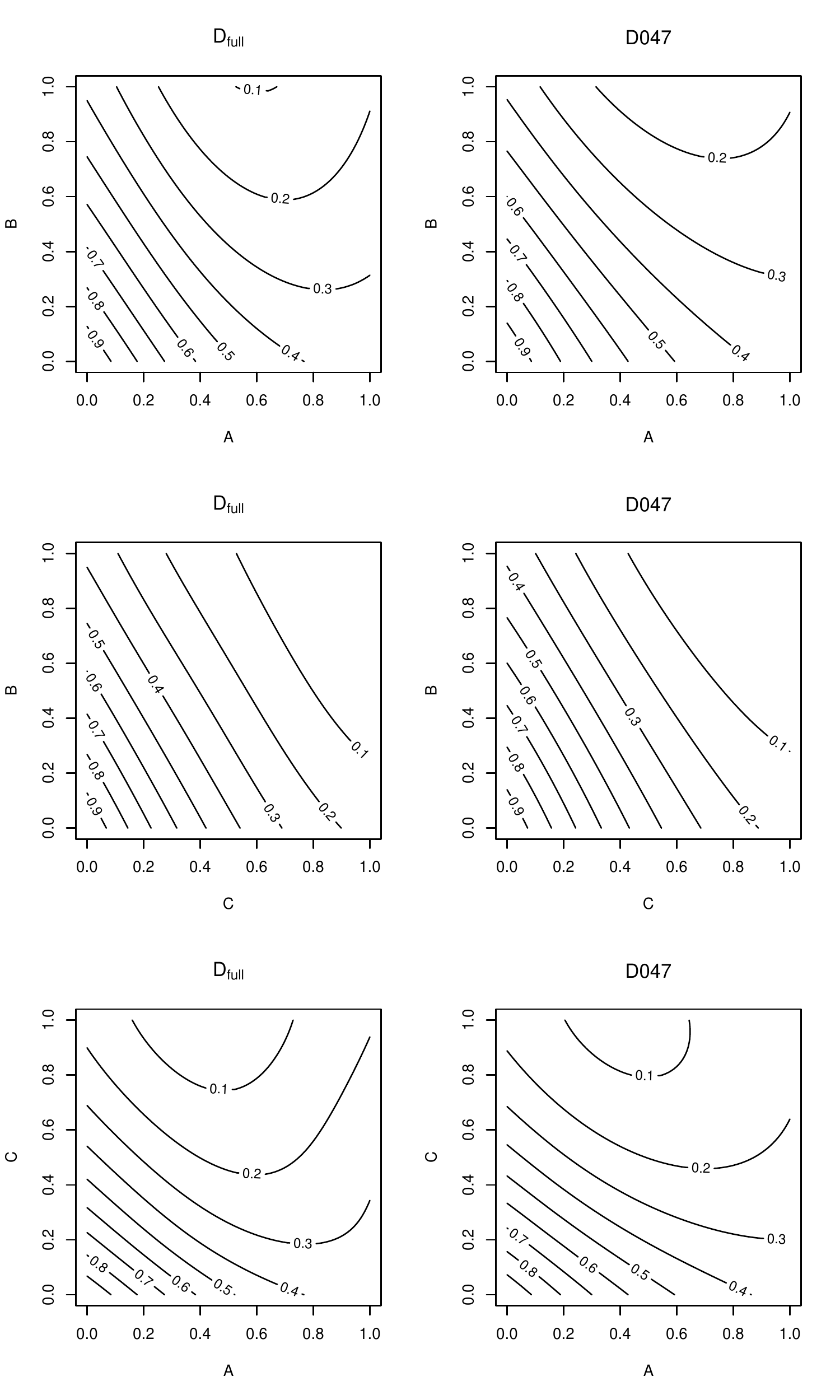}
    
    \label{fig: contournor}
\end{figure}

\begin{figure}
\caption{Contour plots of predicted ATP levels via $D_{\text{full}}$ and $D047$ under Kriging models on  cancer cells.}
    \centering
    \includegraphics[width=0.8\linewidth]{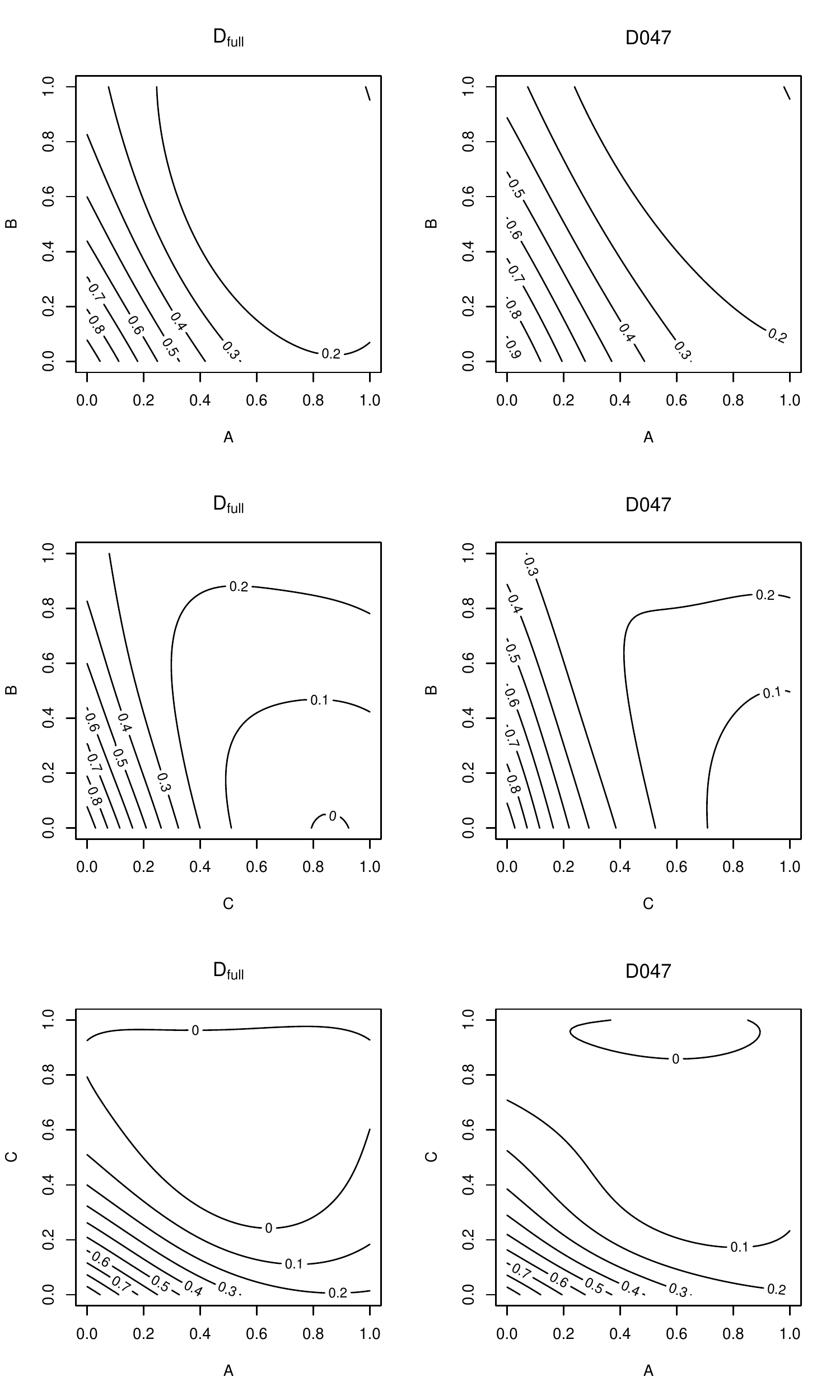}  
    \label{fig: contourcan}
\end{figure}

Figures \ref{fig:nor} and \ref{fig:can} show the scatter-plots of predicted versus observed responses for all four models with design $D047$. From the  figures, we can see that Kriging models are the best in prediction for both normal and cancer cells. Polynomial models perform well for normal cells, but bad for cancer cells; neural networks perform bad for both cases and they require larger designs to achieve accuracy; Hill-based models perform OK, but worse than the Kriging models. 

In order to study the drug interactions, we investigate and compare the contour plots using Kriging models with designs  $D_{\text{full}}$ and $D047$.  Figures \ref{fig: contournor} and \ref{fig: contourcan} show contour plots for any two drugs while fixing the third to 0. We can see that for normal cells, $D_{\text{full}}$ and $D047$ perform nearly the same for all drug combinations; for cancer cells, $D_{\text{full}}$ and  $D047$ perform similarly for A/B and B/C interactions, but slightly different for the A/C interaction. Furthermore, for any two-drug combination, when both dosages are low, the lines are nearly straight, which suggest no interactions; when both dosages are medium, the curves are convex, which suggest synergism; when both dosages are high, curve patterns and their interactions vary by cases. For example, for cancer cells, if we use the highest level of drug C only, the response is 0.011; while, if we use the highest levels of both drugs B and C at the same time, the response is 0.247, which shows clear  antagonism. 
Note that synergism means the two drugs work cooperatively and antagonism means the two drugs inhibit each other.

\section{Discussions}
\label{concl1}

In this paper, we compare four major types of response surface models and four types of designs in analyzing a combinatorial drug experiment by \cite{al2011}. We find that Kriging models need the least number of runs and give the best prediction. Design $D047$ is sufficient in this study, since the response measurement in this experiment is accurate to 2 decimal places and the root MSEs for Kriging models and $D047$ are 1.8\% and 3.3\% for normal and  cancer cells, respectively. It is also shown to be  robust under different models and good enough to analyze two-drug interactions.
Note that if the coded levels (0,4,7) rather than their corresponding actual dosages are used in the design matrix of $D047$, the prediction MSEs are 0.016 and 0.012 for normal and cancer cells, which are much worse than current results. The choice of small designs is not unique. Other 27-run full factorial designs $D057$, $D067$ and 25-run uniform projection design ($UPD_{25}$) can give similar or even better results.



Due to the complexity of underlying biological systems,  a systematic quantification of effects for multiple drugs is challenging, and thus various models should be explored for such experiments. In such situations, space-filling fractional factorial designs are ideal due to their robustness \cite{xiao2015, xiao2017, xiao2017b,  xiao2018}. {Space-filling designs are also ideal for Kriging models, since any unobserved point is close to some design points and so the prediction error is small. An interesting topic for the future research is how space-filling designs perform under Kriging models in drug combination studies.

\bibliographystyle{asa}
\bibliography{ref}

\end{document}